\documentclass[conference]{IEEEtran}
\IEEEoverridecommandlockouts
\usepackage{cite}
\usepackage{amsmath,amssymb,amsfonts}
\usepackage{algorithmic}
\usepackage{graphicx}
\usepackage{textcomp}
\usepackage{xcolor}
\usepackage [english]{babel}
\usepackage [autostyle, english = american]{csquotes}
\MakeOuterQuote{"}
\def\BibTeX{{\rm B\kern-.05em{\sc i\kern-.025em b}\kern-.08em
    T\kern-.1667em\lower.7ex\hbox{E}\kern-.125emX}}
\begin{document}

\title{Deep Learning of Protein Structural Classes:\\ 
Any Evidence for an `{\textit{Urfold}}’?
}

\author{

\IEEEauthorblockN{1\textsuperscript{st} Menuka Jaiswal}
\IEEEauthorblockA{\textit{School of Data Science} \\
\textit{University of Virginia}\\
Charlottesville \\
mj9hre@virginia.edu}
\and
\IEEEauthorblockN{1\textsuperscript{st} Saad Saleem}
\IEEEauthorblockA{\textit{School of Data Science} \\
\textit{University of Virginia}\\
Charlottesville \\
ss3vy@virginia.edu}
\and
\IEEEauthorblockN{1\textsuperscript{st} Yonghyeon Kweon}
\IEEEauthorblockA{\textit{School of Data Science} \\
\textit{University of Virginia}\\
Charlottesville \\
yk4we@virginia.edu}
\and
\IEEEauthorblockN{2\textsuperscript{nd} Eli J Draizen}
\IEEEauthorblockA{\textit{Department of Biomedical Engineering} \\
\textit{University of Virginia}\\
Charlottesville \\
ed4bu@virginia.edu}
\and
\IEEEauthorblockN{2\textsuperscript{nd} Stella Veretnik}
\IEEEauthorblockA{\textit{Department of Biomedical Engineering} \\
\textit{University of Virginia}\\
Charlottesville \\
veretnik@sdsc.edu}
\and
\IEEEauthorblockN{2\textsuperscript{nd} Cameron Mura}
\IEEEauthorblockA{\textit{Department of Biomedical Engineering} \\
\textit{University of Virginia}\\
Charlottesville \\
cmura@virginia.edu}
\and
\IEEEauthorblockN{2\textsuperscript{nd} Philip E Bourne}
%I removed "dean" because it's atypical to put academic titles/positions as part of the name
\IEEEauthorblockA{\textit{School of Data Science} \\
\textit{University of Virginia}\\
Charlottesville \\
peb6a@virginia.edu}
}

\maketitle{ }

\begin{abstract}
Recent computational advances in the accurate prediction of protein three-dimensional (3D) structures from amino acid sequences now present a unique opportunity to decipher the interrelationships between proteins. This task entails—but is not equivalent to—a problem of 3D structure comparison and classification.  Historically, protein domain classification has been a largely manual and subjective activity, relying upon various heuristics. Databases such as CATH represent significant steps towards a more systematic (and automatable) approach, yet there still remains much room for the development of more scalable and quantitative classification methods, grounded in machine learning.  We suspect that re-examining these relationships via a Deep Learning (DL) approach may entail a large-scale restructuring of classification schemes, improved with respect to the interpretability of distant relationships between proteins. Here, we describe our training of DL models on protein domain structures (and their associated physicochemical properties) in order to evaluate classification properties at CATH’s “homologous superfamily” (SF) level. To achieve this, we have devised and applied an extension of image-classification methods and image segmentation techniques, utilizing a convolutional autoencoder model architecture.  Our DL architecture allows models to learn structural features that, in a sense, `define' different homologous SFs.  We evaluate and quantify pairwise `distances' between SFs by building one model per SF and comparing the loss functions of the models. Hierarchical clustering on these distance matrices provides a new view of protein interrelationships—a view that extends beyond simple structural/geometric similarity, and towards the realm of structure/function properties.
\end{abstract}

\section{Introduction}\noindent
Proteins are key biological macromolecules that consist of long, unbranched chains of amino acids (AAs) linked via peptide bonds [1]; they perform most of the physiological functions of cellular life (enzymes, receptors, etc.).  Different sequences of the 20 naturally-occurring AAs can fold into tertiary structures with variable degrees of geometric similarity.  At the sequence level, the differences between any two proteins can range from relatively modest single-point edits ("point mutations" or `substitutions') to larger-scale changes such as reorganization of entire segments of a polypeptide chain.  Such changes are critical because they influence 3D structure, and protein function stems from 3D structure [2]. Indeed, our ability to elucidate protein function and evolution is intimately linked to our knowledge of protein structure.  Equally important, interrelationships between structures define a map of the protein universe [3].  Thus, it is paramount to have a robust classification system for categorically organizing protein structures based upon their similarities. Even more basically, what does `similarity' mean in this context (geometrically, chemically, etc.)?  And, are there particular formulations of `similarity' that are more salient than others?  Ideally, any system of comparison would also take into account functional properties—i.e., not just raw geometric shape, but also properties such as acidity/basicity, physicochemical features of the surface, "binding pockets", and so on.

An unprecedented volume of protein structural and functional data is now available, largely because of exponential growth in genome sequencing and efforts such as structural genomics [4]; alongside these data, novel computational approaches can now discover subtle signals in sets of sequences [5]. These advances have yielded a vast trove of potential protein sequences and structures, but the utility of the information has been limited because the mapping from sequence to structure (or `fold') has remained enigmatic; this decades-long grand challenge is known as the "protein folding problem"[6].  As computational approaches to the folding problem continually improve, increasingly we will compute 3D structures from protein sequences \emph{de novo}. Therefore, we expect to see the demand for identifying and cataloging new protein structures grow at an ever-increasing pace with the rise in the number of 3D structures, both experimentally determined as well as computationally predicted (via physics-based simulations [7] or via Deep Learning approaches [8]).\\
\indent
Historically, efforts to catalog protein structures have been a largely manual and painstaking process, fraught with heuristics. There has been a shift in the paradigm since the introduction of methodical hierarchical database structures, such as CATH, which engender more robust classification schemes into which new protein structures can be incorporated [9]. This is critical as we expand our knowledge of known protein structures. The CATH database has seen phenomenal growth, going from 53M  protein domains classified into 2737 homologous SFs in 2016 [10] to a current 95M  protein domains classified into 6119 SFs. Despite being one of the most comprehensive resources, the CATH database (like any) is not without its limitations, in terms of its underlying assumptions about the relationships between entities. For instance, it was recently argued that there exists an intermediate level of structural granularity that lies between the CATH hierarchy’s architecture (A) and topology (T) strata; dubbed the {\textit{Urfold}} [11], this representational level is thought to capture the phenomenon of "{\emph{architectural similarity despite topological variability}}", as exhibited by the structure/function similarity in a deeply-varying collection of proteins that contain a characteristic small $\beta$-barrel (SBB) domain [12].\\
\indent
With recent advances in computing power, Deep Learning methods have begun to be applied to protein structures, in terms of predictions, similarity assessment and classification [13], [14]. However deep neural networks (DNNs), and specifically 3D CNNs, have not yet seen widespread use with protein structures. This is likely the case because: (1) there is no single, `standard'/canonical orientation across the set of all protein structures [15], which is problematic for CNNs (which are not rotationally invariant), and (2) computational demands to train such models are exorbitant [16].  In this paper, we present a new Deep Learning method to quantify similarities between protein domains using a 3D-CNN based autoencoder architecture.  Our approach treats protein structures as 3D images, with any associated properties (physicochemical, evolutionary, etc.) incorporated as regional attributes (on a voxel-by-voxel basis).  To obviate the problem of angular dependence, we apply random rotations of a given protein; yielding multiple copies of the protein domain, note that these geometric transformations are essentially a form of data augmentation [17].  In this work, we adapted current deep learning architectures, such as the CNNs used in image segmentation and classification tasks [18], for application to our 3D protein classification problem.  A benefit of our approach, as it pertains to proteins, is that 3D protein structures are rather sparse, in terms of the fractional occupancy of voxels in a region of 3D space to which the CNN is applied; this feature can be leveraged for rapid computation via so-called sparse CNNs [19].  Past work with 3D medical images has shown the viability of using sparse CNN architectures for classification and cellular morphological analysis [20].

\section{Methods}

\subsection{Datasets and Initial Featurization }
\noindent
The primary source of data for this project was the CATH protein structure classification database [9]. CATH is a hierarchical classification scheme that organizes all known protein 3D structures (from the Protein Data Bank [PDB; [21]]) by their similarity, with implicit inclusion of some evolutionary information. The PDB houses nearly 180,000 biomolecular structures, determined via experimental means such as X-ray crystallography, nuclear magnetic resonance (NMR) spectroscopy, and cryo-electron microscopy (cryo-EM). CATH uses both automated and manual methods to parse each polypeptide chain in a PDB structure into individual domains. Domain-level entities are then classified within a structural hierarchy: Class, Architecture, Topology and Homologous superfamily (see also Fig 1 in[11] for more on this). If there is compelling evidence that two domains are evolutionarily related (i.e., homologous, based on sequence similarity), then they are classified within the same superfamily. For each domain, we obtain 3D structures and corresponding homologous superfamily labels from CATH.  Next, we compute a host of derived properties for each domain in CATH (Draizen et al., \textit{in prep})---including (i) purely geometric/structural quantities, e.g. secondary structure [21], solvent accessibility [22], (ii) physicochemical properties, e.g. hydrophobicity, partial charges,  electrostatic potentials [23])), and (iii) basic chemical descriptors (atom and residue types). As the initial use-cases that are reported here, we examined three homologous superfamilies of particular interest to us: namely, immunoglobulins (2.60.40.10), the SH3 domain (2.30.30.100), and the OB fold (2.40.50.140). Our models were built using 5583, 2834 and 585 domain structures of Ig, SH3 and OB, respectively.

\subsection{Preprocessing; Further Data Engineering }
\noindent
We began by considering the aforementioned features for each atom in the primary sequence of each domain.  Our first step was to 'clean' the data by eliminating the features that contained more than 25 percent missing values.  Next, we converted continuous-valued numerical features into binary. For example, hydrophobicity values were mapped to 1 if positive and 0 otherwise.  Next, we examined potential correlations among features and eliminated from further consideration any which were redundant (i.e., highly correlated with an existing feature). At this point, our main concern was the computational expense of the problem at hand, and our consideration that it would be cost-ineffective to train convolutional models that incorporate detailed protein structural features which are redundant (e.g., expected training times of several days on a cloud infrastructure). At the end of the preprocessing step, we were left with 38 features that included one-hot encoded representations of (i) atom type (C, CA, N, O, OH), (ii) element type (C\_elem, N\_elem, O\_elem, S\_elem), and (iii) residue type (ALA, CYS, ASP, GLU, PHE, GLY, HIS, ILE, LYS, LEU, MET, ASN, PRO, GLN, ARG, SER, THR, VAL, TRP, TYR).  We also included physicochemical, secondary structural, and residue-associated binary properties at the atomic level (e.g,. is\_hydrophobic, is\_electronegative, positive\_charge, atom\_is\_buried, residue\_is\_buried, is\_helix, is\_sheet).  Next, we represented protein domains as voxels (3D volumetric pixels) using an in-house discretization approach (Draizen et al.; in prep). Briefly, our method centers protein domains in a 256$^3$ Å$^3$ cube (to allow large domains), and each atom is mapped to a 1×1×1 Å$^3$ voxel using a k-d tree data structure (the search ball radius is set to the van der Waal radius of the atom). If two atoms share the space in a given voxel, the maximum between their feature vectors is used because they all contain binary values. Because a significant fraction of voxels do not contain any atoms (proteins are not cube-shaped!), protein domain structures can be encoded via a sparse representation; this substantially mitigates the computational costs of our deep learning workflow.

\subsection{Model Design and Training}
\noindent
A 3D-CNN autoencoder was built for each of the three homologous superfamilies considered here (Ig, SH3, OB).  Our network architecture is inspired by U-Net[18], a convolutional network used in biomedical image segmentation.  The U-Net, which attempts to recreate the input after passing it through the network, consists of a contractive path and an expansive path, giving the eponymous U-shaped architecture.  In the contractive path, each hidden layer contains two 3×3×3 convolutions followed by a rectified linear unit (ReLU), and then a 2×2×2 max pooling layer with strides of two in each dimension.  During the contraction path, the spatial information is reduced (down-sampled), while feature information is increased.  In the expansive path, each layer consists of an up-convolution of 2×2×2, by strides of two in each dimension, followed by two 3×3×3 convolutions and, finally, each of those followed by ReLU nodes. In the convolutional layers, we utilized submanifold sparse convolution operations [24]—an approach that can exploit sparsity of the data (the case for protein domains) in building computationally efficient sparse networks.  The sparse implementation of U-Net replaces the max pooling operation with another convolutional operation.  Our network has 32 filters in the first layer, and double the number of filters each time the data is downsampled; there are five layers of downsampling (Figure 1). We use a linear activation function in the final layer.  The convolutional blocks in our network utilize an approach from the Oxford Robotics Institute’s Visual Geometry Group (VGG); VGG-style blocks have been shown to work well in object recognition [25]. To avoid overfitting, we performed extensive data augmentation and added dropout layers with a rate of 0.5.  For data augmentation, we applied random rotations (mentioned above) to each protein structure; these rotations were in the form of orthogonal transformation matrices drawn from the Haar distribution, which is the uniform distribution on the 3D rotation group (i.e., \textsf{\emph{SO(3)}}; [26]).  In our implementation, we do not concatenate the high resolution features from the contracting path with the upsampled output as done in[18]. 
\begin{figure}[htbp]
\centerline{\includegraphics[width=0.45\textwidth]{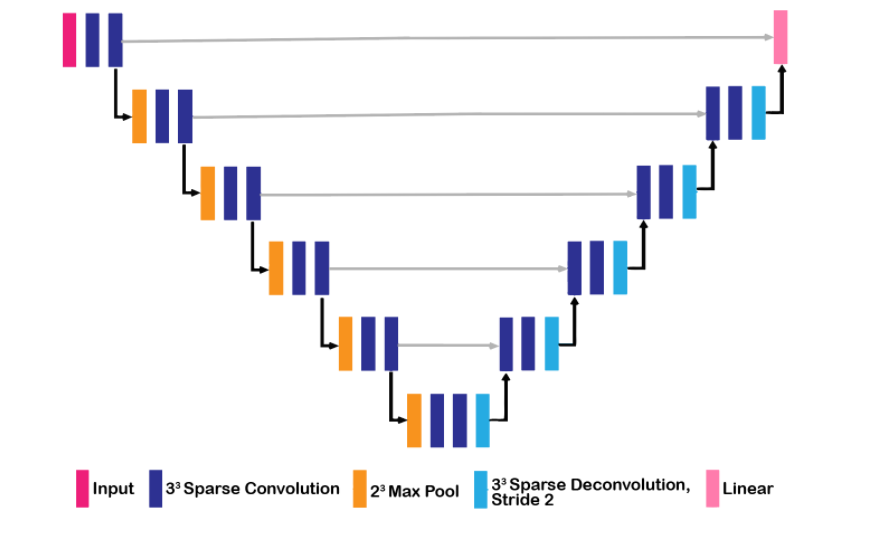}}
\caption{Illustrations of our final network architecture. Dark blue boxes represent sparse convolutions,  orange boxes represent size-2, stride-2 downsampling convolutions; light blue boxes represent the deconvolutions. The greyed out arrows indicate no concatenation of features from the contracting path with the upsampled output.}
\label{fig}
\end{figure}
This ensures that the network does not inadvertently learn to skip the lower network blocks, which would effectively short-circuit itself and contribute to overfitting.

\indent We optimize against sum-squared errors in the output of our model.  We use sum-squared errors because of the relative ease of optimization; however, this may not be ideal for our task of binary classification at the level of each voxel, and this could be a  direction to pursue in future implementations of our model. We used stochastic gradient descent (SGD) as the optimization algorithm, with a momentum of 0.9 and 0.0001 weight decay.  We began with a learning rate of 0.01 and decreased its value by a factor of $e^{((1-epoch)*decay)}$ in each training epoch, using 0.04 as the learning rate decay factor (as suggested in [24]).  Our final network has around 5M parameters in total and all the networks were trained for 100 epochs, using a batch size of 16.  We used the open-source PyTorch framework for all training and inferences [27].

\subsection{Evaluation of Model Performance }
For an autoencoder model with binary input features, the output is expected to be binary as well. Hence, we treat our problem as a binary classification task at the level of each voxel and each feature.  We calculate the area under the Receiver Operating Characteristic curve (AUC) [28] as the primary measure for evaluating the performance of our deep learning models.  The average AUC of the model trained and tested on the Ig superfamily was 0.81, while for SH3 and OB it was 0.88 and 0.89, respectively.  This indicates that the SH3 and OB structures were more readily learned, versus the Ig superfamily structures.

\section{Result and Discussion}
The approach developed and implemented in this work, as illustrated by the initial results described here, can help us validate and otherwise assess existing classification schemes (e.g., CATH, SCOP [29], ECOD [30]). Perhaps most long-term, we believe our methodology can lay a broad foundation for a robust, quantitative, and automatable/scalable mechanism for protein structure classification. This capability, in turn, would represent an advance on many fronts—for example, as a basis for improved processing pipelines for biomolecular structural data and, even more fundamentally, as regards our understanding of biomolecular evolution.  Ultimately, can deep learning help us discover more 'natural' groupings of proteins?  The remainder of this section describes our initial findings, using the Ig, SH3 and OB folds as intriguing use-cases.

\subsection{Feature Importance via Analysis of ROC Curves}\label{AA}\noindent
The determination and extraction of `optimal' features plays a critical role in areas such as protein sequence analysis, as well as in the prediction of protein structures, functions and interactions [31].  Knowing the most `important' features (e.g., those with the greatest predictive power) becomes especially important when there exists abundant data, but there also exist severe limitations in terms of either computational resources or else helpful models/abstractions with which to compute on the data.  Therefore, our initial analyses were concerned with obtaining the most important (predictive) features for our task of protein classification.  As mentioned above, we extracted four categories of features: type of atom, type of amino acid, corresponding physicochemical properties, and secondary structural properties. To evaluate the impact of each feature group on the reconstructbity of our models, we individually utilize each set of features to build the autoencoder and evaluate the area under the ROC curve. Figure 2 shows the ROC curve and AUC values for the top six features; note that the plots in this figure are the averages of independent ROC curves obtained by submitting all protein domains of one variety (e.g., "all Ig domains") into the respective superfamily model (e.g., "the Ig-only model") that was trained on a single feature. This ROC plot reveals that the most important (accuracy-determining) features are (i) atom type, (ii) certain physicochemical properties (burial, electronegativity), and (iii) secondary structural class ({\textsf{is\_sheet}).

\begin{figure}[htbp]
\centerline{\includegraphics[width=0.5\textwidth]{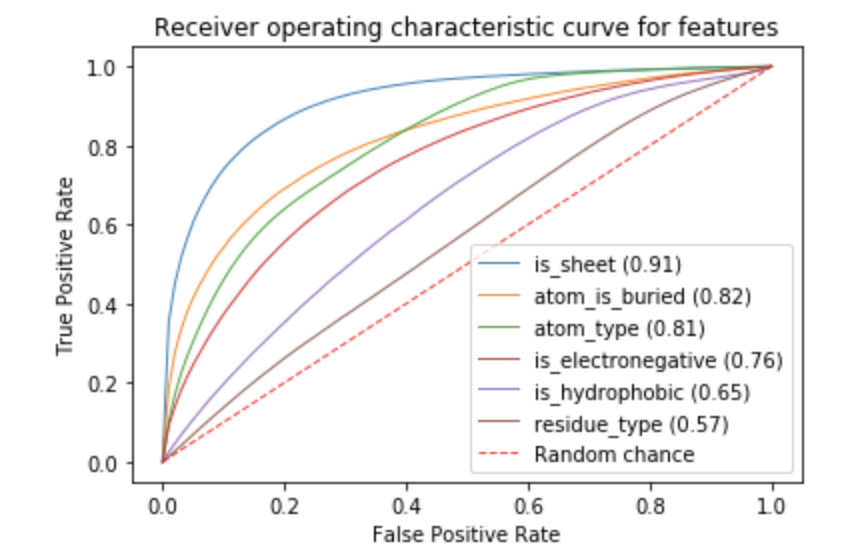}}
\caption{The ROC curve and AUC values for the top six features. ROC curves were obtained by training six models, one model per feature. The values in parentheses besides each feature name are the corresponding AUC values.}
\label{fig}
\end{figure}

\subsection{Reconstruction-based Clustering of Protein Superfamilies}\noindent
To obtain potential clusters of similar protein domains (`similar' under our 3D-CNN/autoencoder approach), we first trained three autoencoder models, one for each of the Ig, SH3 and OB homologous superfamilies.  Next, randomly selected out-of-family proteins were passed into a family-specific model (e.g., a random Ig passed into the SH3-only model), and reconstruction AUC values for each sample were generated.  The basic idea is to utilize the reconstruction AUC as a metric of similarity between (i) a randomly-chosen trial structure, and (ii) the superfamily that is represented (and hopefully accurately `captured') by a given model.  In this way, 150 random samples of domains were selected (50 from each of the SH3, OB and Ig superfamilies), and the reconstruction AUC from the three superfamily models were generated for each of these 150 domains.  Using these reconstruction AUC vectors as features, hierarchical agglomerative clustering was performed using the Python scikit-learn package [32]. The dendrogram in Figure 3 shows the clusters obtained via single-linkage ("nearest neighbor") clustering using the Euclidean distance, or else using Ward’s method, as the parameters in our hierarchical clustering tasks.
The dendrogram clearly indicates the existence of two dominant clusters in the data. The results of cluster assignment remained consistent with the use of \emph{k}-means clustering, with a silhouette score [33] of 0.56. (The silhouette measures intra-cluster `cohesion' versus inter-cluster `separation'; it ranges from -1 [poor grouping] to +1 [good grouping].) Table I is the confusion matrix of superfamily vs obtained cluster labels. As can be seen from the dendrogram and the confusion matrix, the domain clustering that we computed (i.e., {\texttt{\{Ig,\{SH3,OB\}\}}}) differs in some detail from that provided by the CATH classification system.
\begin{figure}[htbp]
\centerline{\includegraphics[width=0.5\textwidth]{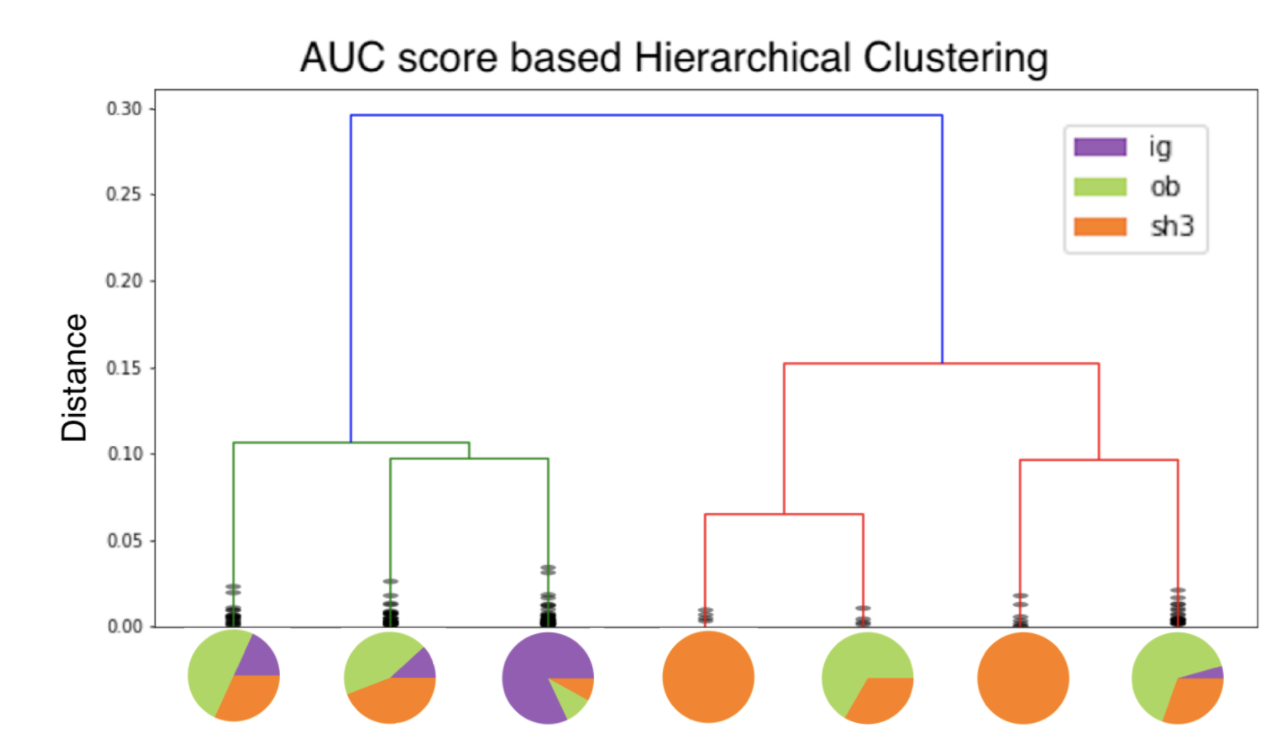}}
\caption{Hierarchical agglomerative clustering based on the AUC scores for various proteins from the Ig, OB and SH3 superfamilies. Note the relatively cohesive grouping of the Ig fold in this dendrogram, while instances of the OB and SH3 folds do not segregate nearly as `cleanly'.}
\label{fig2}
\end{figure}

\begin{table}[htbp]
\caption{}
\begin{center}
\begin{tabular}{ |c|c|c|c| } 
\hline
Superfamily & Cluster1 & Cluster2 \\
\hline
Ig & 1 & 49 \\ 
Sh3 & 19 & 31 \\ 
Ob & 24 & 26 \\ 
\hline
\end{tabular}
\end{center}
\end{table}
In particular, the SH3 and OB superfolds "cross-contaminate" one another, whereas the  majority of Ig domains are cleanly separated into distinct clusters. Our finding that the SH3 and OB do not cleanly cluster (vs the separation exhibited by Ig) is consistent with the recent proposal of an \textit{Urfold} [11] level of protein structure―namely, the notion that there may exist entities that can exhibit similarity at intermediate levels of structural granularity, i.e. between the clear-cut A $\leftrightsquigarrow$ T $\leftrightsquigarrow$ H, etc. levels of CATH’s hierarchical classification scheme.\\
\indent We also examined the average template modeling (TM) score [34] for pairs within the clusters and for pairs within the superfamilies.  In our manual spot-checking of a few cases, we invariably found that the average TM-score computed under our clustering scheme was lower than the average TM-score computed within each of the CATH superfamilies. This is notable, as it indicates that the clusters obtained using the AUC reconstruction-based score are not driven purely by geometric properties of the domains but instead that other properties (e.g., physicochemical) of the domains also influence the goodness of fit to our superfamily model. 

\subsection{Future directions}
To further validate and establish the initial results reported here, our methodology can be applied to broader sets of protein domains, sampled across many more homologous superfamilies (and other levels in the C-A-T-H hierarchy).  Our findings thus far, though limited to only the Ig, SH3 and OB superfamilies, are nevertheless quite interesting: we believe that application of our approach to greater numbers of superfamilies can yield even greater insight into the nature of protein interrelationships (groupings).  To improve the performance of our 3D-CNN-based deep learning models, we suspect that probabilistic frameworks such as variational autoencoders (VAEs; [35]) can be fruitfully employed [36].  VAEs have recently emerged as a powerful approach for unsupervised learning of complicated distributions (i.e., highly intricate input $\rightarrow$ output mappings [latent spaces], with little to no causal information available). For instance, VAEs using concepts from Bayesian machine learning have proven effective in semantic segmentation and visual scene understanding tasks [37], and we anticipate that coupling our 3D-CNN approach to a VAE (versus the simple AE used here) could enhance our protein domain classification methodology. Specific benefits of applying VAEs to our protein problem could include a relaxed need to rely on labelled data (e.g., superfamily labels), and also the capacity to discover relationships between two 'groupings' of proteins, A and B, that have been hitherto entirely unknown—a type of result that would have deep evolutionary implications.

\indent To try and elucidate the basis for the classification decisions of our models, i.e., demystify the typical AI "black box" by making our DNN interpretable, we plan to implenent the layer-wise relevance propagation (LRP) [38] algorithm; this method is prominent in the realm of explainable machine learning [39]. The LRP algorithm—by explicitly tracking the patterns of learned weights (dependencies) from layer to layer (akin to the backpropagation algorithm)—effectively provides a 'picture' of what elements from the input domain map to particular features of the model’s output; this, in turn, can afford immense explanatory power, particularly in the context of a geometric object like a 3D protein structure.  For our model, LRP has the potential to highlight which voxels in the structure of a protein domain contribute to the reconstruction of the output, and to what extent, perhaps extracting the `urfold.'

\section{Conclusion}
In this paper, we proposed a new approach for measuring similarity between protein domains and assessing existing protein classification schemes. We defined an autoencoder model that incorporates residue and constituent atom information of 3D-structures of protein domains as well as their structural and physicochemical properties. We successfully implemented the proposed model and we empirically demonstrated that it allows us to effectively and efficiently learn the defining properties of protein 3D structures, at least at the superfamily level. Specific results from our calculations showed that considering secondary structural and physicochemical information, in addition to pure geometric information, greatly improves our ability to cluster protein domains. Interestingly, one of the most important features identified by our models is the is\_beta\_sheet feature (which defines a major structural element in proteins). Since structural motifs \(\alpha , \beta\) are closely associated with protein fold, this result supports the idea that inclusion of secondary motif information alone in our 3D structural similarity classification task disproportionately improved the performance of our models [40]. From our reconstruction-based clustering, the SH3 and OB superfolds "cross-contaminate" one another, whereas the  majority of Ig domains are cleanly separated into distinct clusters.  Based on the hierarchical classification scheme of CATH, which is predicated largely on 3D structure but also accounts for sequence similarity (H level), superfamilies which belong to the Ig, SH3, and OB folds do indeed differ at the level of architecture (A).  Thus, our finding-- specifically, of SH3 and OB co-clustering implies that there may well exist more important factors, beyond purely geometric and structural similarities, to map the relationship between protein superfamilies; this concept is epitomized by the recent notion of an ‘Urfold’[11] level of structural granularity, consisting of entities between CATH’s ‘A’ and ‘T’ levels. Future directions will further explore this promising finding.

\section*{Acknowledgements}\noindent
We thank Loreto Peter Alonzi for assistance with high-performance computing resources, and Gerard Learmonth and Abigail Flower for feedback and guidance. Portions of this work were supported by the University of Virginia School of Data Science and NSF {\textsc{Career}} award MCB‐1350957.

\vspace{12pt}
\end{document}